\documentclass{elsart}

\usepackage{apalike}
 \usepackage{graphicx}
\usepackage{amssymb}
\usepackage{subfigure}
\usepackage{hyperref}
\begin{document}
\begin{frontmatter}
\title{Co-Authorship Networks in the Digital Library Research Community}

\author[lanl]{Xiaoming Liu},
\author[odu]{Johan Bollen},
\author[odu]{Michael L. Nelson} and 
\author[lanl]{Herbert Van de Sompel}
\address[lanl]{Research Library,
 Los Alamos National Laboratory, \\
Los Alamos, NM, 87545 USA}
\address[odu]{Computer Science Department,
Old Dominion University, \\
Norfolk, VA, 23529 USA}
\begin{abstract}

The field of digital libraries (DLs) coalesced in 1994: the first
digital library conferences were held that year, awareness of the
World Wide Web was accelerating, and the National Science Foundation
awarded \$24 Million (U.S.) for the Digital Library Initiative (DLI).
In this paper we examine the state of the DL domain after a decade of
activity by applying social network analysis to the co-authorship network
of the past ACM, IEEE, and joint ACM/IEEE digital library conferences.
We base our analysis on a common binary undirectional network model to
represent the co-authorship network, and from it we extract several
established network measures.  We also introduce a weighted directional
network model to represent the co-authorship network, for which we define
$AuthorRank$ as an indicator of the impact of an individual author in
the network.  The results are validated
against conference program committee members in the same period. The
results show clear advantages of PageRank and AuthorRank over degree,
closeness and betweenness centrality metrics.  We also investigate the
amount and nature of international participation in Joint Conference on Digital Libraries (JCDL).

\end{abstract}
\end{frontmatter}

\section{Introduction and Motivation}
In 1994, the National Science Foundation awarded \$24 Million (U.S.) to
six institutions, thereby officially kicking off the federally-sponsored
DL research program. Also in 1994, the first of what was later to become
the IEEE Advances in Digital Libraries (ADL) conference and the ACM
Digital Libraries (DL) conference were held in New Jersey and Texas,
respectively. In 2001, the two conference series were merged and the
first ACM/IEEE Joint Conference on Digital Libraries (JCDL) was held
in Virginia.  These conferences have induced a pattern of collaborations
which has shaped the domain of DLs over the past decade.  
To study the structure of these collaborations, and thereby learn more about
the DL research community itself, we used social network analysis to
investigate authorship trends in the composite corpus of the DL, ADL
and JCDL conferences.

Many co-authorship networks have been studied
\cite{newman1,sigir,farakas2002:physicaa,cuuningham1997:scientometrics,leo:jasis}
to investigate the structure of scientific collaborations, and several have studied 
DL discipline in general \cite{Mutschke:01,cunningham}. The DL community
offers an interesting case study for a number of reasons.  Firstly,
it is a quickly growing, dynamic field which has only existed since
approximately 1994.  Investigations of its present status and structure
will yield valuable data for future longitudinal studies. Secondly,
the domain of DLs is a highly multidisciplinary community which has
attracted researchers from a wide area of expertise, e.g.~databases,
networking, information and library science, human computer interaction,
high performance computing, archiving, and education.
 This enriches DL research with the expertise of a variety of scholars,
but may lead to fractionating of the community.  Lastly, in such a dynamic,
and new domain, few journals exist that are peer-reviewed
and included in the ISI Journal Citation Reports.  This makes it difficult
to assess the status, impact and influence of researchers and their
institutions if traditional methods cannot be applied.

We are interested in the structure of collaborations within the DL
research community and quantitative metrics for the concepts of status
and influence.  In this paper, we study author status by determining
author centrality in a co-authorship network derived from the ADL,
DL and JCDL conferences from 1994-2004.  Other DL conferences exist:
the European Conference on Digital Libraries (ECDL) began in 1997,
the International Conference on Asian Digital Libraries (ICADL) began
in 1998, and the Russian Conference on Digital Libraries (RCDL) began
in 1999.  In addition to these conferences, the DL research community
is covered by online serials such as D-Lib Magazine and the Journal of
Digital Information.  Although there is a Journal of Digital Libraries,
much of the DL research results are covered in traditional journals by
the respective communities outlined above.   We chose the ADL, DL and
JCDL conference series because of our familiarity with the conferences,
the ease of automated data collection of them, their longevity, their
sponsorship by the ACM and IEEE, and the fact that they were the first such
conferences to be held.  Although ADL, DL and JCDL are international
conferences, the fact that they are always held in the U.S. will surely
influence the results, because attendance of an author is required for
paper acceptance.

To perform this analysis, we built a weighted directional network model
to represent collaboration  relationships.  We applied a variety
of centrality measures to investigate this network and then defined
AuthorRank, an alternative centrality metric which exploits the features
of such networks.  The result is validated against the set of past DL, ADL
and JCDL program committee members on the assumption that program committee
members can be regarded as prestigious actors in a field.  Our results
show clear advantages of the use of AuthorRank and PageRank.
\section{Background and Related Work}
Social network analysis has attracted considerable interest
in recent years and plays an important role in many disciplines
\cite{otte,wasserman,jscott,barabasi,watts}.   A popular culture example
is the Oracle Of Bacon project \cite{oracleofbacon}, which determines
the distance between any actor and Kevin Bacon by examining movie
co-starring relationships.  This fun example demonstrates the usefulness
that can arise by adapting the concept of a relationship in social network
analysis to the domain of interest.  By defining a relationship to be the
co-authoring of an ADL, DL or JCDL conference paper, we can bring social
network analysis methods to bear on our analysis of the DL research
community.  
\subsection{Social Network Analysis}
Social network analysis is based on the premise that the relationships
between social actors can be described by a graph. The graph's nodes
represent social actors and the graph's edges connect pairs of nodes
and thus represent social interactions.  This representation allows
researchers to apply graph theory \cite{wasserman} to the analysis of
what would otherwise be considered an inherently elusive and poorly
understood problem: the tangled web of our social interactions. In
this article, we will assume such graph representation and use the
terms \emph{node}, \emph{actor},  and \emph{author} interchangeably.
The terms \emph{edge}, \emph{relationship}, and \emph{co-authorship}
are also used interchangeably.

Given that we have established a social network graph, we can describe
its properties on two levels, namely by global graph metrics and
individual actor properties.  Global graph metrics seek to describe the
characteristic of a social network as a whole, for example the graph's
diameter, mean node distance, the number of components (fully connected
subgraphs), cliques, clusters, small-worldness, etc.  Actor properties
relate to the analysis of the individual properties of network actors,
e.g.~actor status, distance, and position in a cluster.

The status of an actor is usually expressed in terms of its
centrality, i.e.~a measure of how central the actor is to the network
graph. Central actors are well connected to other actors and metrics
of centrality will therefore attempt to measure an actor's degree
(number of in- and out- links), average distance to all other actors,
or the degree to which geodesic paths between any pair of actors passes
through the actor.

A class of impact metrics focuses on the recursive nature of
status. Clearly, when one is endorsed by a high status actor,
this increases one's status more than being endorsed by a low status
actor. Hence, one's status can be derived from the status of the actors
one is linked to. This leads to a recursive definition of status which
is mathematically addressed by eigenvector analysis. Since the web's
hyperlink structure mimics the properties of a social network graph
(WWW pages are nodes, hyperlink are edges), eigenvector analysis can also
used to measure the prestige of web pages; well-known algorithms include
PageRank \cite{page98pagerank}, SALSA \cite{lempel00stochastic} and HITS
\cite{kleinberg99authoritative}. However, in these algorithms all edges
by definition have binary weights: a hyperlink either exists or does not exist,
 and a social relationship exists or does not exist.  Bharat and
Henzinger \cite{bharat98improved} developed a weighted edge scheme to
improve the HITS algorithm. Given its formulation, it is also possible to
modify the assumption of equiprobability underlying PageRank's formulation
 to take edge weight into account \cite{chakrabarti:mining}.

\subsection{Co-Authorship Networks}
Co-authorship networks are an important class of social networks and
have been used extensively to determine the structure of scientific
collaborations and the status of individual researchers.  Although
somewhat similar to the much studied citation networks in the scientific
literature \cite{garfield}, co-authorship implies a much stronger social
bond than citation.  Citations can occur without the authors
knowing each other and can span across time.  Co-authorship implies a
temporal and collegial relationship that places it more squarely in
the realm of social network analysis. 

An early example of a co-authorship network is the Erd\"{o}s
Number Project, in which the smallest number of co-authorship links
between any individual mathematician and the Hungarian mathematician
Erd\"{o}s are calculated \cite{decastro99famous}.  (A mathematician's
``Erd\"{o}s Number'' is analogous to an actor's ``Bacon Number''.)
Newman studied and compared the co-authorship graph of arXiv, Medline,
SPIRES, and NCSTRL \cite{newman1,newman2} and found a number of
network differences between experimental and theoretical disciplines.
Co-authorship analysis has also been applied to various ACM conferences:
Information Retrieval (SIGIR) \cite{sigir}, Management of Data (SIGMOD)
\cite{sigmod} and Hypertext \cite{carr99hypertext}, as well as mathematics
and neuroscience \cite{farakas2002:physicaa}, information systems
\cite{cuuningham1997:scientometrics}, and the field of social network analysis \cite{otte}.
 International co-authorship networks
have been studied in Journal of American Society for Information Science
\& Technology (JASIST) \cite{compar:he2002}  and Science Citation Index
\cite{mappin:wagner2004}.

\section{Constructing Co-Authorship Networks}
We present the representational foundations of our work by discussing three approaches to 
model co-authorship networks. The
first model is a traditional undirected, binary graph, the second model is a
directed, binary network which allows calculation of actor prestige, and
in the third model we consider weighted co-authorship relations in the network. A
set of centrality and prestige metrics is adapted to operate on the
resulting graphs. In particular, we propose AuthorRank, a weighted
version of PageRank.

\subsection{Binary, Undirected Co-Authorship Networks}

A simple and widely used co-authorship network model is based on an undirected, binary
graph $G$ in with each edge represents a co-authorship relationship.

Consider two articles:

\begin{tabular}{ccc}
article 	&			&	authors\\\hline
article 1 	& $\rightarrow$	&	$\{v_1, v_2, v_3\}$\\
article 2	& $\rightarrow$	&	$\{v_1, v_2\}$\\
\end{tabular}
\vspace{1cm}

\begin{center}
\begin{figure}%[htp]
\centering
\subfigure[Binary undirected network]{\label{fig:simple}\includegraphics{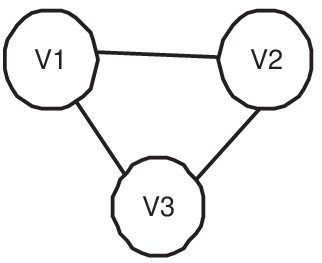}}\quad\quad\quad\quad
\subfigure[Binary directed network]{\label{fig:directed}\includegraphics{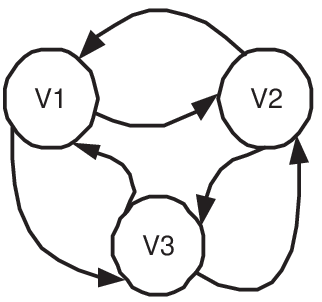}}
\vspace{.5in}
\subfigure[Weighted directed network]{\label{fig:weighted}\includegraphics{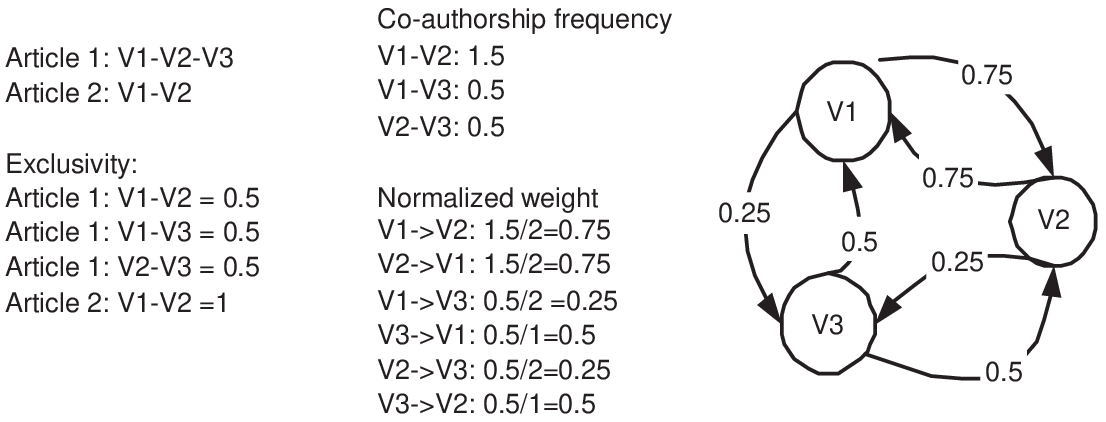}}
\caption{Representations of co-authorship network}
\label{fig:binarynetwork}
\end{figure}
\end{center}

If any two authors co-authored an article, an edge with unit weight is
created (Figure \ref{fig:simple}). For example, in the table above,
authors $v_1$ and $v_2$ would be connected by an edge since they
co-authored article 1.

The resulting graph is denoted as an undirected unit-weighted graph
$G=(V,E)$, where the set of $n$ authors is denoted $V=\{v_1,...v_n\}$
and $E \subseteq V^2$ represents the edges between authors. As will be
shown in following sections, various graph metrics can be extracted from
this kind of network.

\subsection{Binary, Directed Co-Authorship Networks}

In order to measure \emph{prestige} of an author, we must distinguish
``endorsement'' accorded from endorsement received by authors. In social
network analysis, the concept of prestige is defined for directional
relationships. In order to convert a co-authorship graph to a directed
graph, we make the following assumptions:

\begin{enumerate}
	\item any undirected network can be represented as a directed network with symmetric
		linkage, i.e. every edge in the undirected network $G$ is replaced by two, symmetrical
		directed edges;
	\item the resulting directional, symmetrical edges represent the mutual endorsement of authors.
		In fact, in a random walk model, the directional edges can be understood as the bi-directional movement
		of a surfer;
	\item The edge weight is a binary value, indicating the presence or absence of two symmetrical edges.
\end{enumerate}

The resulting graph is denoted as a directed unit-weighted graph (Figure
\ref{fig:directed}). As will be shown in the following sections, PageRank
and other prestige measures can be applied to this network.

\subsection{Weighted, Directed Co-Authorship Networks}

The binary graph representation of co-authorship
network omits a number of factors which shape collaboration patterns
among authors.  There are many cases in which the binary
network does not correspond with a common sense notion of magnitude. For
example, if two authors co-publish many papers, should the link between
them be considered more important than the link between occasional
co-authors? Also, if one article has two authors and another article has
a hundred authors, should the authors in the first article be considered
more connected than those of the second article? 

To allow an expression of relationship magnitude we represent the
co-authorship network as a directed weighted graph. The co-authorship
graph $G$ is denoted $G=(V,E,W)$, where $V$ is the set of nodes (authors),
$E$ is the set of edges (co-author relationships between authors), and
$W$ is the set of weights $w_{ij}$ associated with each edge connecting
a pair of authors $(v_i,v_j)$.

We propose to determine the magnitude of the link between two authors
on the basis of two factors:

\begin{enumerate}
	\item Frequency of co-authorship: authors that frequently co-author should have a higher co-authorship weight.
	\item Total number of co-authors on articles: if an article has many authors, each individual co-author
		relationship should be weighted less.
\end{enumerate}

We can now determine the weight of co-authorship links.  Let the set
of $n$ authors be denoted as $V=\{v_1,...v_n\}$. Let the set of $m$
articles be denoted as $A=\{a_1,...,a_k,...a_m\}$, and $f(a_k)$ be the
number of authors of article $a_k$. We define:

\textbf{Exclusivity: } If authors $v_i$ and $v_j$ are co-authors in article $a_k$,
\begin{equation}
\label{equ1}
g_{i,j,k}=1/(f(a_k)-1)
\end{equation}

$g_{i,j,k}$ represents the degree to which author $v_i$ and $v_j$
have an exclusive co-authorship relation for a particular article.
This definition gives more weight to co-author relationships in
articles with fewer total co-authors than articles with large numbers
of co-authors, i.e.~it weighs the co-authorship relation in terms of
how exclusive it is.

\textbf{Co-authorship frequency:} \begin{equation}
c_{ij}=\sum\limits_{k=1}^{m}{g_{i,j,k}} \end{equation} The co-authorship
frequency consists of the sum of all $g_{i,j,k}$ values for all articles
co-authored by  $v_i$ and $v_j$.  This gives more weight to authors who
co-publish more papers together, and do so exclusively.

\textbf{Normalized weight:}
\begin{equation}
w_{ij}=\frac{c_{ij}}{\sum\limits_{k=1}^{n}{c_{ik}}}
\end{equation}
This normalization ensures that the weights of an author's relationships sum to one.

The notions of exclusivity and frequency used in determining
co-authorship relations correspond to the principles underlying Term
Frequency vs.  Inverse Document Frequency (TFIDF) weighting used in IR
\cite{modern:baezayates1999}. A TFIDF term weight expresses how strongly
a term is tied to a particular document on the basis of how frequently the
term occurs in the document itself versus how frequently it occurs in all
documents in the collection. In other words, a term which is exclusively
tied to a particular document will be most frequent within the document
itself, i.e. its term frequency is high, while being relatively rare
across the collection, i.e. its document frequency is low. In the same
manner, we normalize the raw co-authorship frequency by the number of
co-authors, the latter an indication of how exclusive or non-exclusive
the co-authorship relations is.

The proposed weighting scheme also has an intuitive basis in random
walks on graphs (Figure \ref{fig:weighted}).  The normalized weight
corresponds to the probability distribution of a random walk on the
co-authorship graph.  A random walker may choose to start navigating
the network from any author. In Figure \ref{fig:weighted}, if the walk
starts from author $v_1$, the walker may travel to $v_2$ or $v_3$ with
probability $0.75$ and $0.25$ respectively. If the walker starts from
author $v_3$, however, the walker has the same probability of visiting
$v_1$ or $v_2$.  The weighted co-authorship also has an intuitive
meaning as the endorsement of an author. For example, from Figure
\ref{fig:weighted}, we can understand that $v_1$ and $v_2$ have a higher
mutual endorsement since they co-authored more papers.

\subsection{Metrics for Co-Authorship Network}

A number of social network metrics are available for measuring
the characteristics of a binary undirected collaboration network,
including  components analysis, small world analysis, and centrality
analysis. These metrics measure various network properties and some
may only be applied under certain conditions.  The metrics 
used in this paper and their applicability  are listed in Table \ref{tbl:measurement} and
discussed below.

\begin{table*}[htbp]
\begin{center}
\caption{Co-authorship network metrics}
\begin{tabular}{|l|c|c|c|c|c|c|c|c|}
\hline
Metric&
\multicolumn{2}{|c|}{Type}& 
\multicolumn{2}{|c|}{Property}& 
\multicolumn{2}{|c|}{Scope} &
\multicolumn{2}{|c|}{Importance}\\
\hline
&  Binary &  Weighted &  Actor& Global&Whole&Largest&Centrality&Prestige\\
&&&&&Network&Component&&\\
\hline
Component&$\times$&&&$\times$&$\times$&&&\\
\hline
Small World&$\times$&&&$\times$&&$\times$&&\\
\hline
Cluster&&$\times$&&$\times$&&$\times$&&\\
\hline
Closeness&   $\times$&&$\times$&&&$\times$&$\times$&\\
\hline
Betweenness&  $\times$&&$\times$&&$\times$&&$\times$&\\
\hline
Degree&    $\times$&&$\times$&&$\times$&&$\times$&\\
\hline
PageRank&    $\times$&&$\times$&&$\times$&&&$\times$\\
\hline
AuthorRank&&$\times$&$\times$&&$\times$&&&$\times$\\
\hline
\end{tabular}
\label{tbl:measurement}
\end{center}
\end{table*}

\subsubsection{Component size analysis}

A component of a graph is a subset with the characteristic that
there is a path between any node and any other node of this subset.
A co-authorship network usually consists of many disconnected components
(e.g. disconnected research groups or individuals), and component analysis
can be used to learn about the structure of the network. Some network
analysis methods are only widely used in connected networks. Therefore,
in networks with disconnected components, those methods are typically
only applied to the largest connected component, as shown in Table
\ref{tbl:measurement}.

\subsubsection{Degree, closeness, betweenness centrality}
We have adapted three common centrality metrics, namely degree centrality,
closeness centrality, and betweenness centrality \cite{wasserman}, for their use on
binary, undirected co-authorship networks.

Degree centrality of a node is defined as the total number of edges
that are adjacent to this node. Degree centrality represents the simplest
instantiation of the notion of centrality since it measures only how many
connections tie authors to their immediate neighbors in the network.

However, authors may be well connected to their immediate neighbors but be
part of a relatively isolated clique. Although locally well connected,
overall centrality is low. Closeness centrality therefore expands
the definition of degree centrality by focusing on how close an author
is to all other authors. To calculate a node's closeness centrality we
determine its shortest-path distances to all authors in the network and
invert these values to a metric of closeness. A central author is thus
characterized by many, short connections to other authors in the networks.

Betweenness centrality represents a different operationalization of
centrality. It is based on determining how often a particular node is
found on the shortest path between any pair of nodes in the network. Nodes
that are often on the shortest-path between other nodes are deemed highly
central because they control the flow of information in the network.
Betweenness centrality can be used in disconnected networks, however it
may generate a large number of nodes with zero centrality, since many
nodes may not act as a bridge in the network.

Though the discussed centrality metrics can be extended to
directed and weighted networks, this has received less attention
\cite{newman-weighted,wasserman}.  In this article we will focus on
their usage in binary, undirected networks.

\subsubsection{Eigenvector centrality or PageRank}
PageRank is the ranking mechanism at the heart of Google
\cite{page,page98pagerank}. In PageRank, a hyperlink is understood as an
``endorsement'' relationship. PageRank's definition of prestige deviates from
the degree, closeness and betweenness centrality by modeling inherited
or transferred status.

A page has high rank if the sum of the ranks of its backlinks is
high. This covers both the case when a page has many backlinks and when
a page has a few highly ranked backlinks.  PageRank can be calculated
using a simple iterative algorithm, and corresponds to the principal
eigenvector of the normalized link matrix of the web. 

PageRank is originally designed to rank retrieval results based on the hyperlink structure
 of the web, which is a directed but binary graph in nature, therefore we apply PageRank 
to the binary directed network model. Our work is inspired by a variety of proposals to 
extend PageRank to weighted and bi-directional networks.  Eigenvalue centrality
was originally intended for an undirected graph \cite{bonacich1}.  Applying
PageRank and related centrality measures in a weighted environment is
discussed in \cite{newman-weighted,newman-betweeness}.  Other variations
and improvements to PageRank include a ``topic sensitive'' PageRank to
improve search performance \cite{topic-pagerank}, distributed
computation techniques for calculating PageRank \cite{p2p-pagerank,distributed-pagerank}, and 
faster computation of PageRank
\cite{extrapolation-pagerank}.

\subsubsection{AuthorRank: PageRank for weighted, directional networks}

%The degree, betweenness, and closeness centralities are not well defined in weighed network.

We submit that PageRank can be applied to an
undirectional co-authorship graph by transforming each undirectional edge
into a set of two directional, symmetrical edges. However, the reduction
of edge weights to binary values entails a severe loss of information. The
generated co-authorship weights express valuable information which should,
and can, be taken into account when calculating PageRank values over a
weighted co-authorship graph.

%PageRank is also defined in a binary network.

\begin{center}
\begin{figure}[tbhp]
\centerline{
\subfigure[PageRank: A connects to B,C,D and transfers 1/3]{\includegraphics []{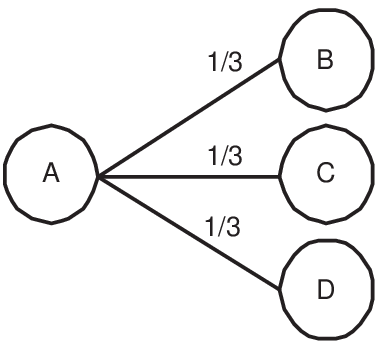}}\quad\quad\quad
\subfigure[AuthorRank: A connects to B,C,D and transfers according to link weight]{\includegraphics[]{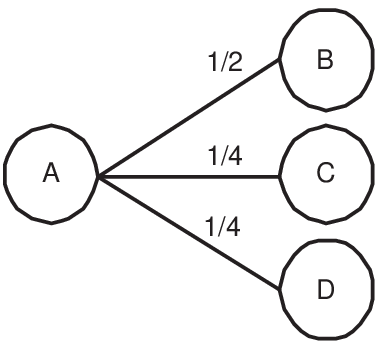}}
}
\caption{Weight transfer in PageRank and AuthorRank}
\label{fig:weighttransfer}
\end{figure}
\end{center}

We therefore define \emph{AuthorRank}, a modification of PageRank which
considers link weight. It is based on a modification of the PageRank
assumption that a node transfers its PageRank values evenly to all the
nodes it connects to. Indeed, PageRank assumes that when a node A connects
to $n$ other nodes, each receives a fraction $\frac{1}{n}$ of $PR(A)$. In
probabilistic terms, this models a random walker who is equally probable
to walk from node A to each of its connecting nodes. However, in reality,
the chances of link traversal can be expected to be distributed quite
unevenly and according to the degree of relationship between $A$ and the
nodes it connects to. Our co-authorship link weights express how strongly
related two nodes, or authors, are in the co-authorship graph and these
weights can therefore be used to determine the amount of PageRank that
should be transferred from node A to the nodes it connects to (Figure \ref{fig:weighttransfer}).

The AuthorRank of an
author $i$ is then given as follows:

\begin{eqnarray*}
AR(i) = (1-d) &	+  &	d \left( AR(1) \times w_{1,i} + \cdots + AR(n) \times w_{n,i} \right)\\
AR(i) = (1-d) &	+  &	d \sum_{j=0}^n AR(j) \times  w_{j,i}
\label{PR}
\end{eqnarray*}

where $AR(j)$ corresponds to the AuthorRank of the backlinking node, and $w_{j,i}$ corresponds to the
edge weight between node $j$ and $i$. The AuthorRank can be calculated with the same iterative algorithm used
by PageRank. One may think of AuthorRank as a generalization of PageRank
by substituting $w_{j,i}$ with $\frac{1}{C(j)}$ in PageRank, in which $C(j)$ is defined as the number      of links going out of page j.

Looking at the example network underlying Figure \ref{fig:binarynetwork} and
Figure \ref{fig:weighttransfer}, AuthorRank better reveals status of actors
than centrality measures and PageRank. When collaboration frequency and exclusivity are considered, 
$v_1$ and $v_2$ are more prestigious than $v_3$ in the network, 
AuthorRank captures this property, while centrality measures and PageRank cannot.

\section{DL Research Community Co-Authorship Analysis}
\subsection{Generating the Co-authorship Network}

We extracted co-authorship data from DBLP (http://dblp.uni-trier.de/) for ACM DL
(1995-2000), IEEE ADL (1994-2000), and JCDL (2001-2003). This includes
all long papers, short papers, posters, demonstrations, and organizers
of workshops. \footnote{Unfortunately, due to an error in DBLP, the DL 94
dataset was omitted.  We do not believe this omission will significantly
alter our findings.} The dataset contained 1567 authors, 759 publications,
and 3401 co-authorship relationship pairs. Some statistics are readily
available from this data set. For example, the number of articles,
authors, international (non-US) authors, and new authors per year are
shown in Figure \ref{fig:year}.  It can be seen that number of articles
and the number of authors are highly correlated, and that a major
boost occurred following the merger of the  ACM/IEEE DL series into a
single JCDL conference.  Figure \ref{fig:paper_author} shows the number
of publications per author.  The values range between 1 and 22, with 4
authors publishing more than 10 papers and 78\% of the authors publishing
only 1 paper and  95\% authors having 3 papers or less.  Authors with 8
or more publications are shown in Table \ref{tbl:publications}. Each paper
has a mean of 3.02 authors and a median of 3 authors. The distribution of
number of authors per paper is shown in Table \ref{tbl:authorsperpaper}.

\begin{center}
\begin{figure}[tbhp]
\centerline{\includegraphics{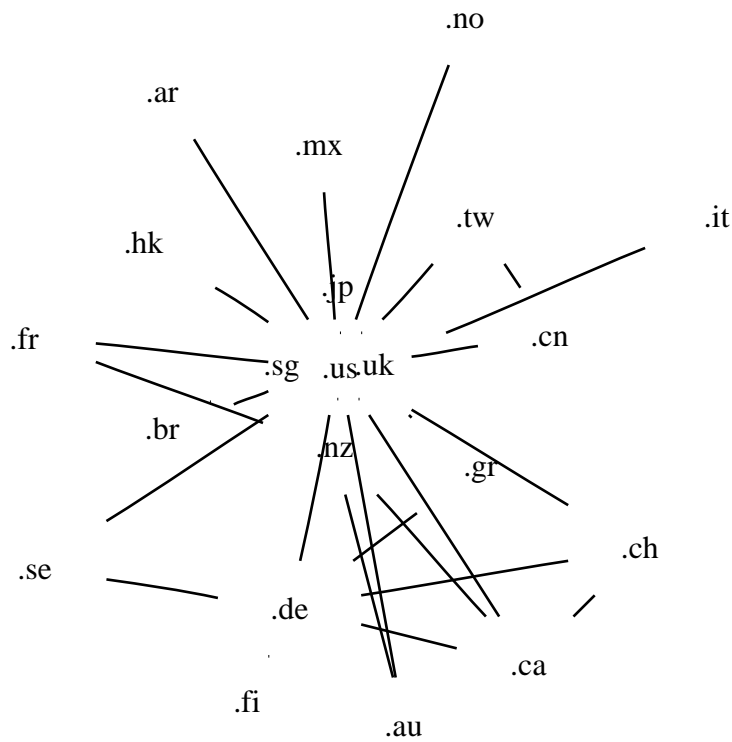}}
\caption{Country network}
\label{fig:ctynetwork}
\end{figure}
\end{center}

\begin{center}
\begin{figure}[tbhp]
\centerline{\includegraphics{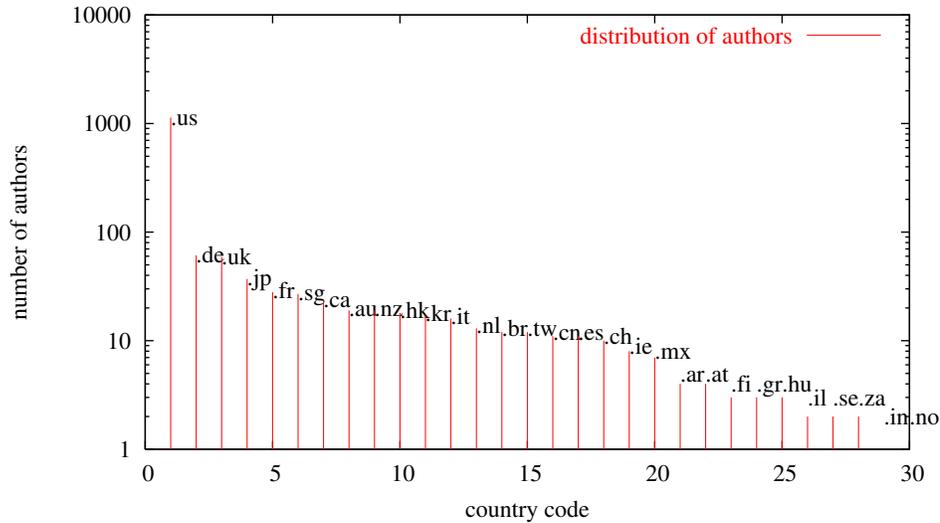}}
\caption{Distribution of authors per country}
\label{fig:country_author}
\end{figure}
\end{center}

We also studied international collaboration.  Approximately 72\%
(1133/1567) of the authors are affiliated with U.S. institutions.
We discovered that among 3401 co-authorship relationships, only about
7\% are collaborations between authors from different countries. A
country collaboration network is created by accumulating cross-country
collaborations from the author network. Figure \ref{fig:ctynetwork} shows the result;
countries are represented by domain names, and two countries are closer to each
other if authors from those countries collaborated closely.  The figure
can only be considered approximate due to the limitations of the
visualization technology used. Figure \ref{fig:ctynetwork} shows that
JCDL community is centered around .us,  with .uk, .nz, and .sg closely
surrounding .us; .nz and .de also play significant roles in connecting
different countries.  There are nine countries (.es, .ie, .at, .hu,
.nl, .in, .kr, .il, and .za; with 61 authors) that are not connected
with other countries. The distribution of authors from each country is
shown in Figure \ref{fig:country_author}.

\begin{center}
\begin{figure}[tbhp]
\centerline{\includegraphics{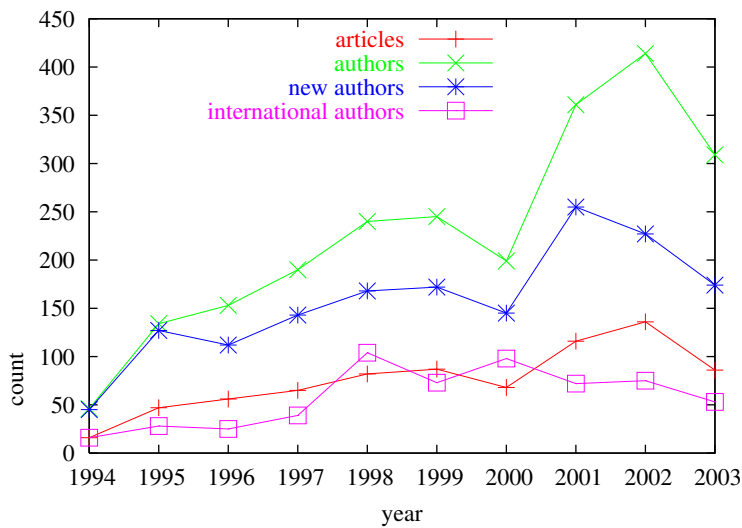}}
\caption{Articles, authors, international authors, and new authors per year}
\label{fig:year}
\end{figure}
\begin{figure}[tbhp]
\centerline{\includegraphics{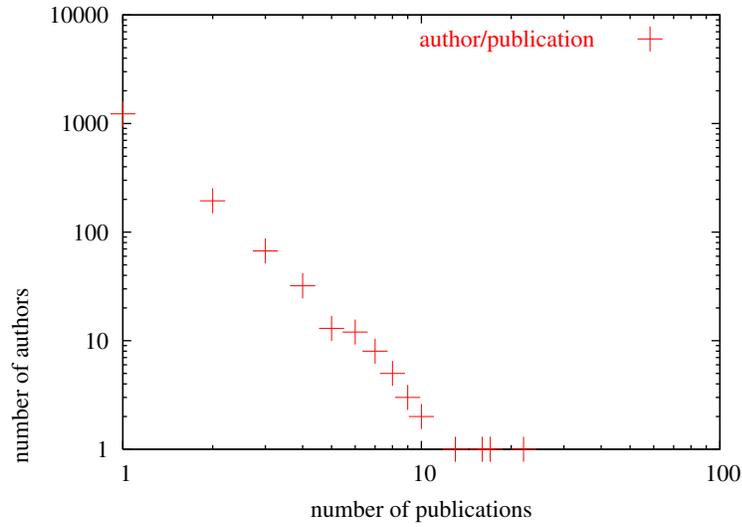}}
\caption{Number of papers per author}
\label{fig:paper_author}
\end{figure}
\end{center}

\begin{table}[tbhp]
\begin{center}
\caption{Authors with 8 or more publications}
\begin{tabular}{|l|r|}
\hline
Name&Publications \\
\hline
\hline
Hsinchun Chen & 22\\
\hline
Edward A. Fox& 17\\\hline
Ian H. Witten& 16\\\hline
Hector Garcia-Molina& 13\\\hline
Alexander G. Hauptmann &10\\\hline
Gary Marchionini& 10\\\hline
Judith Klavans &9\\\hline
Carl Lagoze &9\\\hline
Michael L. Nelson& 9\\\hline
David Bainbridge& 8\\\hline
Richard Furuta& 8\\\hline
Ee-Peng Lim &8\\\hline
Catherine C. Marshall& 8\\\hline
Terence R. Smith &8\\\hline
\end{tabular}
\label{tbl:publications}
\end{center}
\end{table}
\begin{table}[tbhp]
\begin{center}
\caption{Distribution of number of authors per paper}
\begin{tabular}{|l|r|r|r|r|r|r|r|r|}
\hline Number of authors& Number of papers & Percentage\\\hline
1&149&19.6\%\\\hline
2&216&28.5\%\\\hline
3&179&23.6\%\\\hline
4&94&12.4\%\\\hline
5&45&5.9\%\\\hline
6&33&4.3\%\\\hline
7&20&2.6\%\\\hline
8&7&0.9\%\\\hline
9&4&0.5\%\\\hline
10&5&0.7\%\\\hline
11&1&0.1\%\\\hline
12&2&0.3\%\\\hline
13&1&0.1\%\\\hline
14&1&0.1\%\\\hline
15&2&0.3\%\\\hline
total&&100\%\\\hline
\end{tabular}
\label{tbl:authorsperpaper}
\end{center}
\end{table}
\begin{center}
\begin{figure}[tbhp]
\centerline{\includegraphics [width=3.3in]{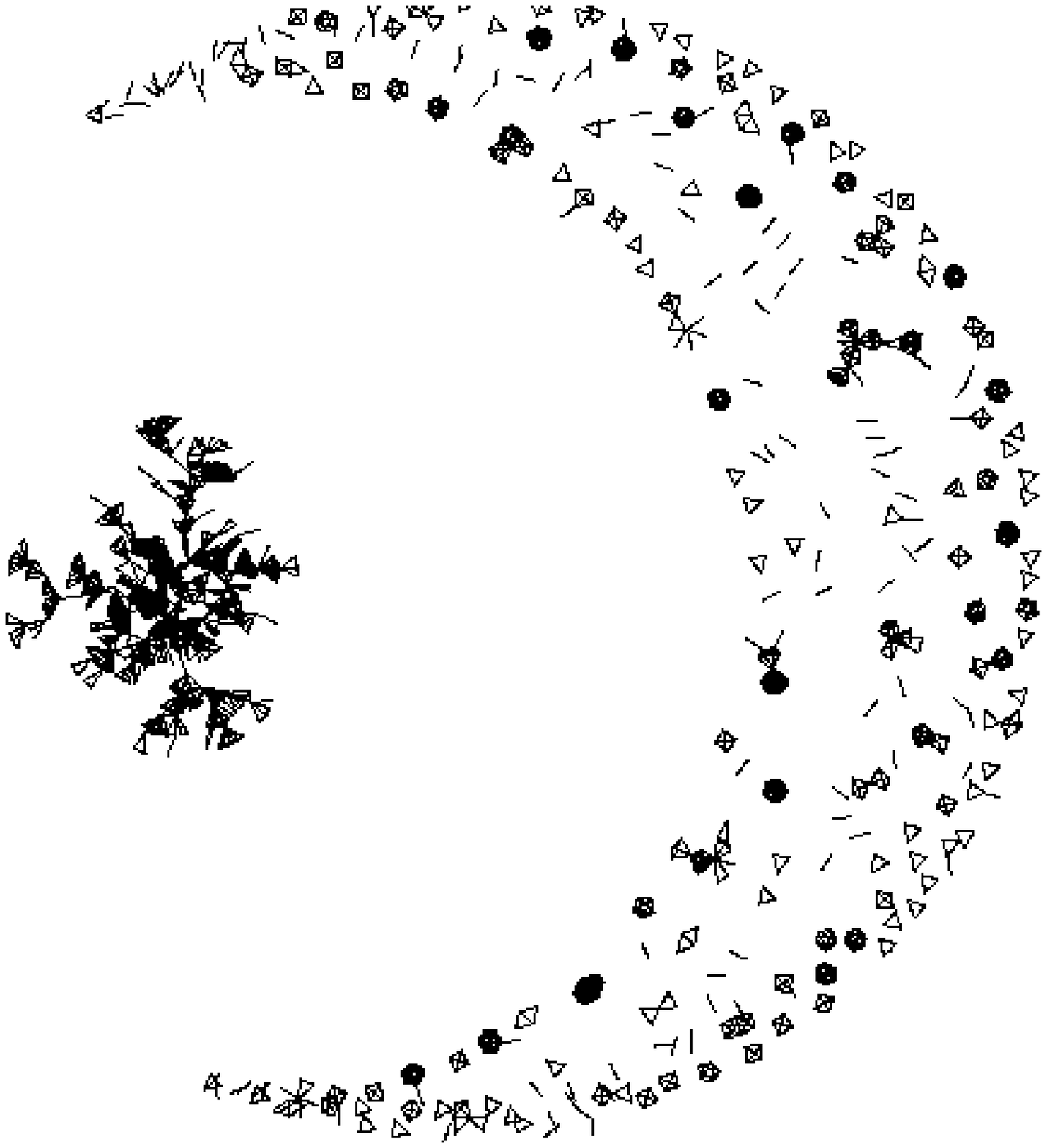}}
\caption{Component size analysis}
\label{fig:com}
\end{figure}
\end{center}

\subsection{Component Size Analysis}

Similar to observations from previous research in co-authorship networks,
the DL co-authorship network is not a single connected graph. The largest
component of the network has 599 authors, the second largest component has
31 nodes and so on.  The entire co-authorship network with all components
is visualized in Figure \ref{fig:com}, in which nodes represent authors
and links represent co-authorship relationship. The largest component is on the left
side of the Figure, while the right side shows many small components.
Well-connected components are recognizable by their very dense (dark)
shape.

Nascimento \cite{sigmod} reports that the largest component in
SIGMOD's co-authorship graph has about 60\% of all authors. In the four
co-authorship networks studied by Newman \cite{newman1}, NCSTRL has the
smallest largest component, containing 57.2\% of all authors. However,
in the JCDL co-authorship network the largest component only includes
38\% of all authors. Several possible explanations could account for
this low value, including the relative immaturity of the the DL field,
the multi-disciplinary nature of the composite JCDL conference series, the
fact that many DL projects grow from a ``grass-roots'', institutionally
oriented focus \cite{esler:jasis1998}, or limited international 
collaboration in the DL research community.

To better understand the nature of major components and the reason for
them not being in the large component, we conducted a manual analysis
of other large components.  This showed that the most dense shapes
include authors from the same institution or working on the same project.
We counted 18 components with sizes ranging from 7 to 31.  By checking the
affiliation of authors, we discovered that 5 components consist mainly of
non-US participants, and that the 31-node component represents
the medical informatics community. By checking titles and content,
we found that 13 components account for short papers or posters only,
many of which are about a specific DL application in a particular scenario.
Therefore, it is our guess that the short paper and poster programs
encourage a wide participation from other disciplines.

\subsection{Cluster Analysis}

The weighted graph model also improves the clustering because close and
frequent collaboration causes higher similarity scores between authors,
resulting in them being grouped closer together. By representing each
author as a vector of relationships to other authors using the weighted
graph model, we conducted a bottom-up, hierarchical clustering algorithm
on the largest component of the co-authorship network. The hierarchical
clustering algorithm starts with all authors and successively combines
them into groups with high inter-authorship similarity. Typically,
the earlier mergers happen between groups with a large similarity, and
similarity becomes lower and lower for later merges. The result reveals that
initial clusters do not necessarily reflect institutional boundaries. This
may be due to the fact that authors may change institutions,  and in
some cases strong collaborations exist between institutions. In the next
stage, institutions are merged into larger clusters due to their joint
publications or common research interests. A well-connected author is
usually only clustered in this stage, which confirms that well-connected
authors play an important role in connecting different clusters.

\begin{center}
\begin{figure}[tbhp]
\centerline{\includegraphics[width=5in]{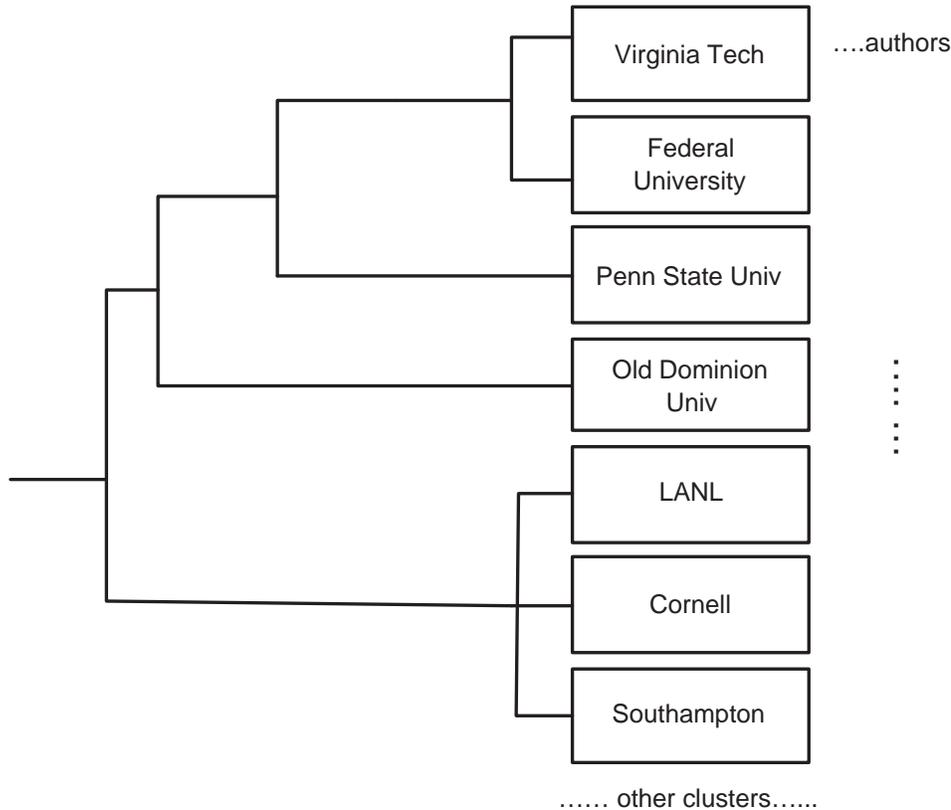}}
\caption{Clustering Result }
\label{fig:cluster}
\end{figure}
\end{center}

As a matter of illustration, the clusters to which the authors of this
paper belong are shown in Figure \ref{fig:cluster}.  As can be seen, small
clusters are initially formed in each authors' institution (Los Alamos National Laboratory and Old
Dominion University), and later institutions are merged to larger clusters.
The frequency of joint publications may explain the different stage of
merging. By checking publications in each cluster, we found that LANL,
Cornell University and the University of Southampton form a larger
cluster because Cornell cooperated with Southampton in the Open Citation
project, and LANL worked with Cornell on the Open Archives Initiative.
Similarly, Virginia Tech (VT) collaborated with the Federal University of Minas Gerais in Brazil in the Web-DL project \cite{827201}, with Penn State (PSU) in the CITIDEL project,
and with Old Dominion University (ODU) in the NCSTRL project. ODU and
PSU have no joint publications, they are clustered together because both
collaborated with VT. VT and Federal University of Minas Gerais  probably merged earlier because they have more joint publications.
\subsection{Small World Analysis}
\begin{center}
\begin{figure}[tbhp]
\centerline{\includegraphics[width=3.3in]{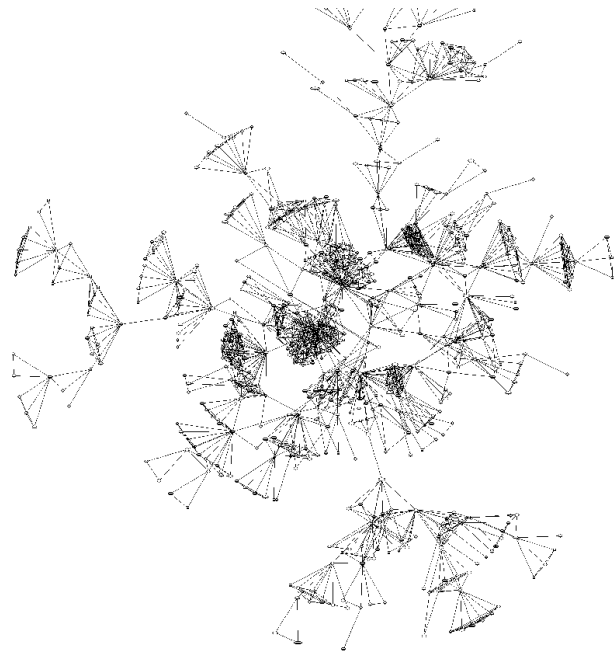}}
\caption{Largest component}
\label{fig:lgcom}
\end{figure}
\end{center}

Since small world analysis can only be done in a connected graph, we used
the largest component of the co-authorship network for our calculation.
The largest component (599 authors and 1897 links) has a clustering
coefficient of 0.89, and a characteristic path length of 6.58.  With a
similarly sized connected random graph, the clustering coefficient is
0.31 and the characteristic path length is 3.66.  This means that the JCDL
co-authorship network is a small world graph as can be expected. The giant component
is shown in Figure \ref{fig:lgcom}.

Nascimento \cite{sigmod} reports that the SIGMOD co-authorship graph
yields a clustering coefficient of 0.69, and a characteristic path
length of 5.65. In all four networks studied by Newman, the largest
clustering coefficient generated is 0.726. This shows a rather high
clustering coefficient of the JCDL co-authorship network, meaning that
co-authors of one author are more likely to publish together. The JCDL
co-authorship network also has a rather long characteristic path length,
indicating that authors from different groups are not as well connected
as, for example, those in the SIGMOD co-authorship network.

\begin{center}
\begin{figure}[tbhp]
\centerline{\includegraphics{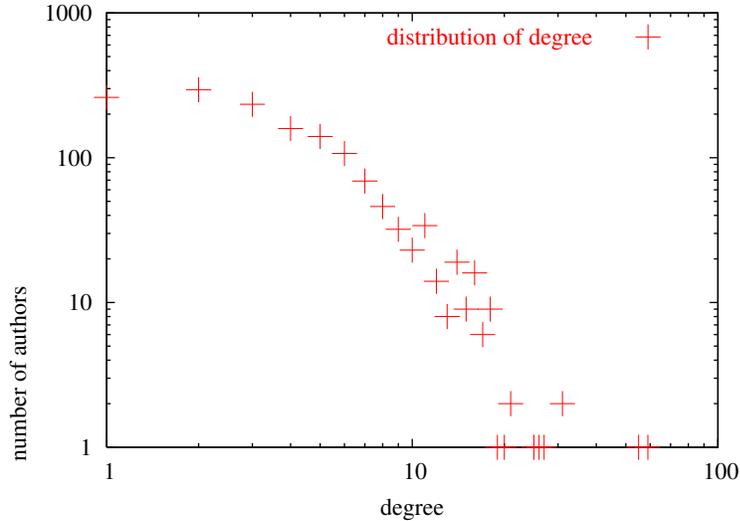}}
\caption{Degree distribution}
\label{fig:degree}
\end{figure}
\end{center}

\subsection{Centrality}
Using the R package (http://www.r-project.org/), we calculated the degree, closeness, and
betweenness centrality for the binary undirected co-authorship network only,
as these metrics are not well defined in a weighted network.  The highest
ranking 20 authors for each metric and their scores are listed in Table
\ref{tbl:centrality}.

\begin{table*}[htbp]
\begin{center}
\caption{Authors ranked according to centrality measure}
\begin{tabular}{|l|l|r|l|r|l|r|}
\hline
rank& 
\multicolumn{2}{|c|}{Degree}& 
\multicolumn{2}{|c|}{Betweenness}& 
\multicolumn{2}{|c|}{Closeness} \\
\hline
1& 
Hsinchun Chen &59& 
Hsinchun Chen &89250.92& 
Hsinchun Chen &0.259 \\
\hline
2& 
Edward A. Fox &55& 
Edward A. Fox &83163.92& 
Edward A. Fox &0.251 \\
\hline
3& 
Terence R. Smith &31& 
Judith Klavans &57422.69& 
Judith Klavans &0.235 \\
\hline
4& 
Carl Lagoze &31& 
William Y. Arms &52242.27& 
Gary Marchionini &0.234 \\
\hline
5& 
Judith Klavans &27& 
Nina Wacholder &39226.5& 
Michael L. Nelson &0.229 \\
\hline
6& 
Zan Huang &26& 
Craig Nevill-Manning &38808.08& 
Yiwen Zhang &0.226 \\
\hline
7& 
Gary Marchionini &25& 
David M. Levy &35769.0& 
Ann M. Lally &0.226 \\
\hline
8& 
William Y. Arms& 21& 
Ann P. Bishop &32280.0& 
Lillian N. Cassel& 0.226 \\
\hline
9& 
Richard Furuta &21& 
Tobun D. Ng &30197.13& 
Byron Marshall &0.225 \\
\hline
10& 
Luis Gravano &20& 
Gary Marchionini &29593.86& 
Rao Shen &0.225 \\
\hline
11& 
Michael Freeston& 19& 
Alexander Hauptmann & 29142.0& 
William Y. Arms & 0.224 \\
\hline
12& 
Ian H. Witten &18& 
Catherine C. Marshall& 28587.0& 
Anne Craig &0.221 \\
\hline
13& 
Hector Garcia-Molina& 18& 
Terence R. Smith& 23691.87& 
Larry Brandt& 0.221 \\
\hline
14& 
Michael G. Christel &18& 
Carl Lagoze &22192.66& 
Terence R. Smith &0.219 \\
\hline
15& 
David Millman &18& 
David Bainbridge &21168.03& 
Tobun D. Ng &0.219 \\
\hline
16& 
Tamara Sumner &18& 
Michael L. Nelson &20696.41& 
James C. French &0.219 \\
\hline
17& 
Diane Hillmann &18& 
Howard D. Wactlar &17577.0& 
Kurt Maly &0.212 \\
\hline
18& 
Yilu Zhou &18& 
Ching-chih Chen& 17309.67& 
Mohammad Zubair &0.212 \\
\hline
19& 
Jialun Qin &18& 
John J. Leggett& 15845.5& 
Hesham Anan &0.212 \\
\hline
20& 
Mary Tiles& 18& 
Elizabeth D. Liddy& 14964.0& 
Xiaoming Liu &0.212 \\
\hline
\end{tabular}
\label{tbl:centrality}
\end{center}
\end{table*}

\subsubsection{Degree centrality}

The degree centrality distribution is shown in Figure \ref{fig:degree}. It
follows a rough power-law distribution with a few authors having
a high degree of connection, and most authors have low degree. This
measurement has the disadvantage of giving many authors the same
weight. It is also biased to authors with many co-authors on a single
publication, which is common in experimental sciences \cite{newman1}. The time complexity
is $O(1)$.

\subsubsection{Closeness centrality}

The closeness centrality is only applied to the largest component (599
authors) since closeness is not well defined in a disconnected network. It
has a bias toward authors that are directly connected to a well-connected
author. For example, we discovered in Table \ref{tbl:centrality} that
graduate assistants of a prestigious professor may have a fairly high
weight. The time complexity is $O(n^2)$, where $n$ is the number of
authors in the network.

\subsubsection{Betweenness centrality}

The betweenness centrality is applied to the whole network, however
only 153 authors have positive values. The remaining 1414 authors do
not lie on the shortest paths between other authors. Betweenness is,
in some sense, a measure of the influence a node has over the spread of
information through the network, and indeed some high-ranking authors
play crucial rules in connecting different communities.

The computation of betweenness
centrality is the most resource-intensive of all measures we explored,
since it requires enumerating all of the shortest paths between each
pair of nodes. The time complexity is $O(n^3)$, where $n$ is the number 
of authors in the network, thus limiting its feasibility in large networks.

\subsection{PageRank and AuthorRank}

We developed a Java program with a MySQL backend to calculate PageRank and
AuthorRank. Both calculations can be completed in several seconds. The
20 highest scoring authors for the PageRank and AuthorRank metrics are
listed in Table \ref{tbl:authorrank}. The time complexity of both algorithms is $O(n)$, where $n$ is the number of authors in the network.

\begin{table}[htbp]
\caption{Authors ranked according to PageRank/AuthorRank}
\begin{center}
\begin{tabular}{|l|l|l|}
\hline
Rank& PageRank & AuthorRank  \\
\hline
1& 
Edward A. Fox& 
Hsinchun Chen \\
\hline
2& 
Hsinchun Chen&  
Edward A. Fox\\
\hline
3& 
Carl Lagoze& 
Ian H. Witten\\
\hline
4& 
Judith Klavans& 
Gary Marchionini\\
\hline
5& 
Richard Furuta& 
Hector Garcia-Molina\\
\hline
6& 
Gary Marchionini& 
Carl Lagoze\\
\hline
7& Michael G. Christel&
Alexander G. Hauptmann\\
\hline
8& 
Terence R. Smith& 
Judith Klavans\\
\hline
9& 
Tamara Sumner&  
Richard Furuta\\
\hline
10& 
Ian H. Witten& 
Terence R. Smith\\
\hline
11& Alexander G. Hauptmann&
Tamara Sumner\\
\hline
12& 
Hector Garcia-Molina& 
Ee-Peng Lim\\
\hline
13& 
Javed Mostafa& 
Michael G. Christel\\
\hline
14& 
Alexa T. McCray& 
Michael L. Nelson \\
\hline
15& 
Ee-Peng Lim& 
Wee Keong Ng\\
\hline
16& 
David Bainbridge& 
Javed Mostafa\\
\hline
17& 
Sally Jo Cunningham& 
David Bainbridge\\
\hline
18& 
Luis Gravano& 
J. Alfredo S\'{a}nchez\\
\hline
19& 
Catherine C. Marshall&
Alexa T. McCray\\
\hline
20& 
W. Bruce Croft& 
Andreas Paepcke\\
\hline
\end{tabular}
\label{tbl:authorrank}
\end{center}
\end{table}

\subsection{Correlation and validation}

Several articles have compared the performance of centrality and prestige
metrics, and a general conclusion can be that no single measure is
suited for all applications; each method has its virtues and utility
\cite{wasserman,chakrabarti:mining}.  We verified and compared
metrics in two ways: by the computation of the Spearman correlation
coefficient across ranking methods, and by cross-validation against the
dataset of JCDL program committee members.
\subsubsection{Spearman Correlation}
The Spearman correlation coefficient is used to measure the strength
of association between two variables.  In our case, since betweenness
generated only 153 authors with positive ranking, and closeness
centrality has only been calculated for the largest component, we only
compare degree centrality, PageRank, and AuthorRank. The correlation
coefficient between the degree centrality and PageRank is 0.52, and the
correlation coefficient between the degree centrality and AuthorRank is
0.30 (Figure \ref{fig:scatterplot}). As expected, PageRank and AuthorRank
are more closely correlated with a correlation coefficient of 0.75
(Figure \ref{fig:scatterplot}).

\subsubsection{Program committee validation}

We also verified each ranking method against a dataset consisting of all
members of the JCDL, ADL and DL program committees from 1994 to 2004. This
is meaningful, as program committee members are assumed to be prestigious
actors in the co-authorship network. To that end, the names of all JCDL,
ADL and DL program committee members were collected from the conference
web sites or printed proceedings.  The highest scoring 50 authors
for each ranking method (degree, closeness, betweenness, PageRank,
AuthorRank) were then matched one by one against each JCDL committee member
to identify matches.

Figure \ref{fig:committee} shows the result of this comparison.  The
highest ranking 5 authors for each metric have an almost perfect match
against the dataset of JCDL program committee members.  Overall closeness
ranking performs the worst, as only six authors of the 50 highest ranking
authors are on the JCDL committees.   This is not a surprise since
closeness measures the distance to other authors, and since an author
next to a prominent author is not necessarily also a prominent author.
Degree centrality had mediocre performance.  Betweenness centrality
performs the best among the three centrality measures.  Since betweenness
evaluates one's importance as a bridge between others, this suggests
a committee member may be more likely to serve as a bridge between
research groups than a non-committee member.  Betweenness, PageRank,
and AuthorRank all show good results, however PageRank and AuthorRank
are feasible in large networks due to their low computational complexity.
The results of PageRank and AuthorRank are highly correlated, but there
is no conclusive evidence that one performs better than the other.

\begin{center}
\begin{figure}[tbhp]
\subfigure[Degree centrality vs. AuthorRank]{\includegraphics[width=3.3in]{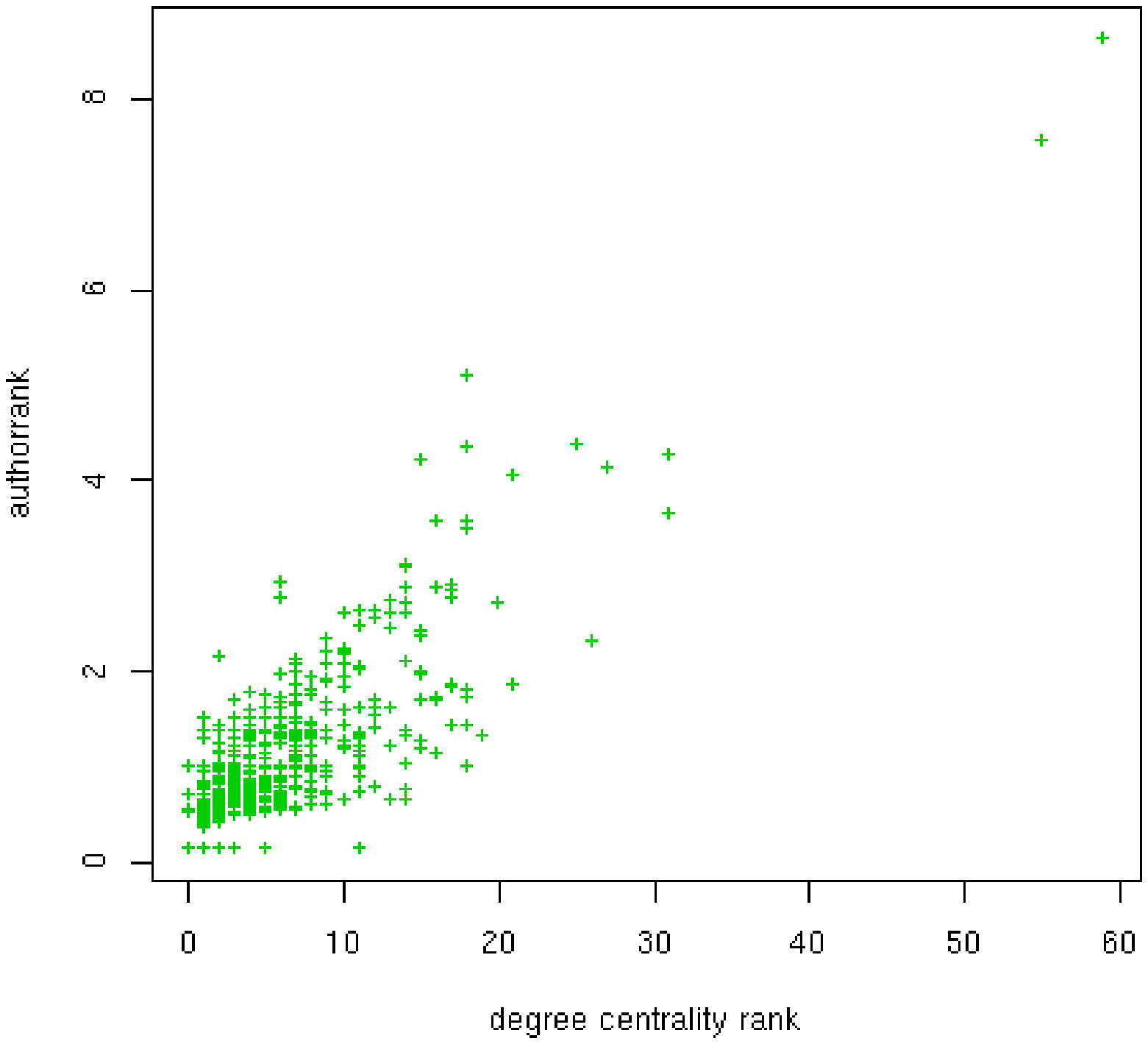}}
\subfigure[PageRank vs. AuthorRank]{\includegraphics[width=3.3in]{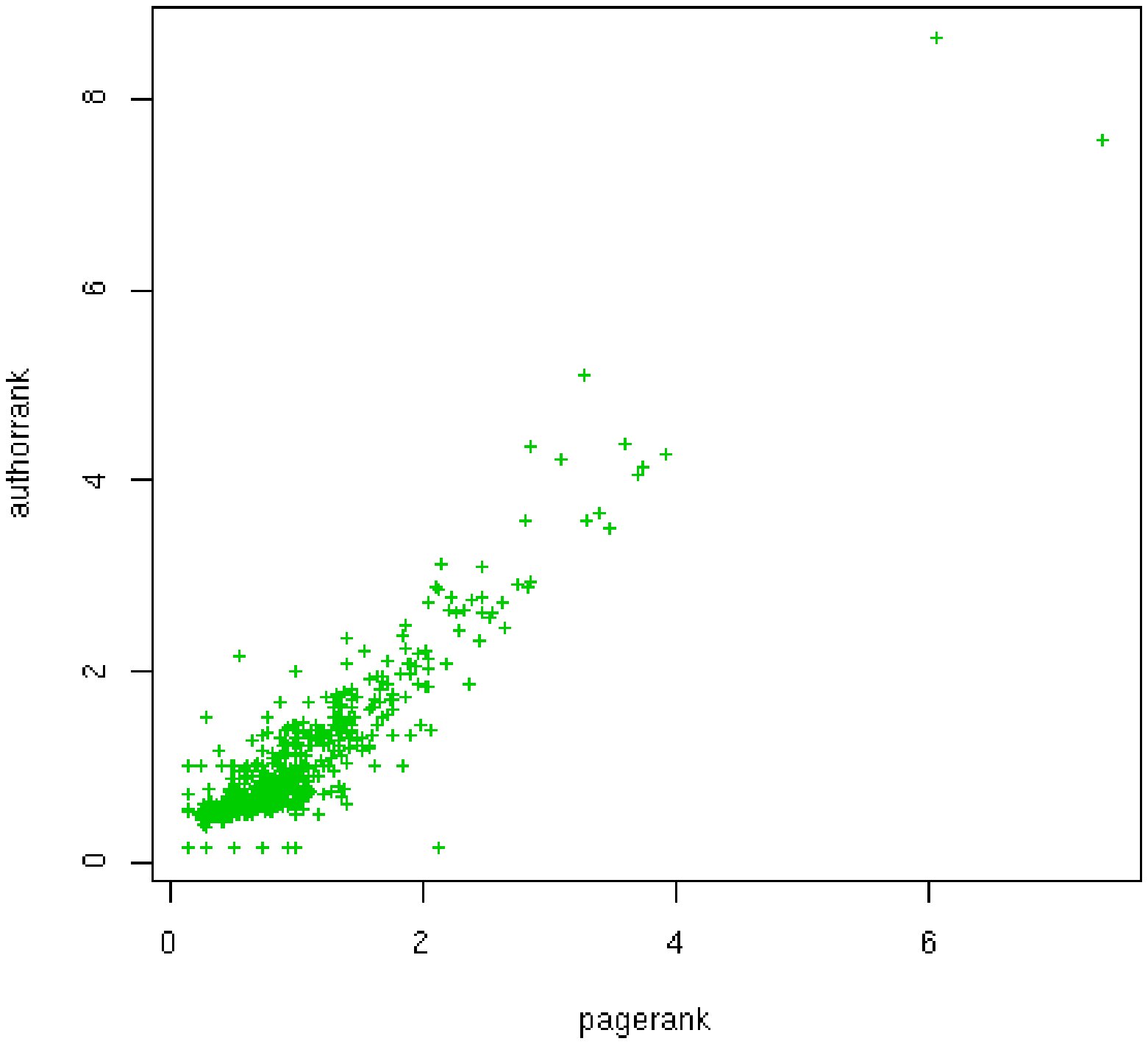}}
\caption{Comparison of ranking algorithms}
\label{fig:scatterplot}
\end{figure}

\begin{figure}[tbhp]
\centerline{\includegraphics{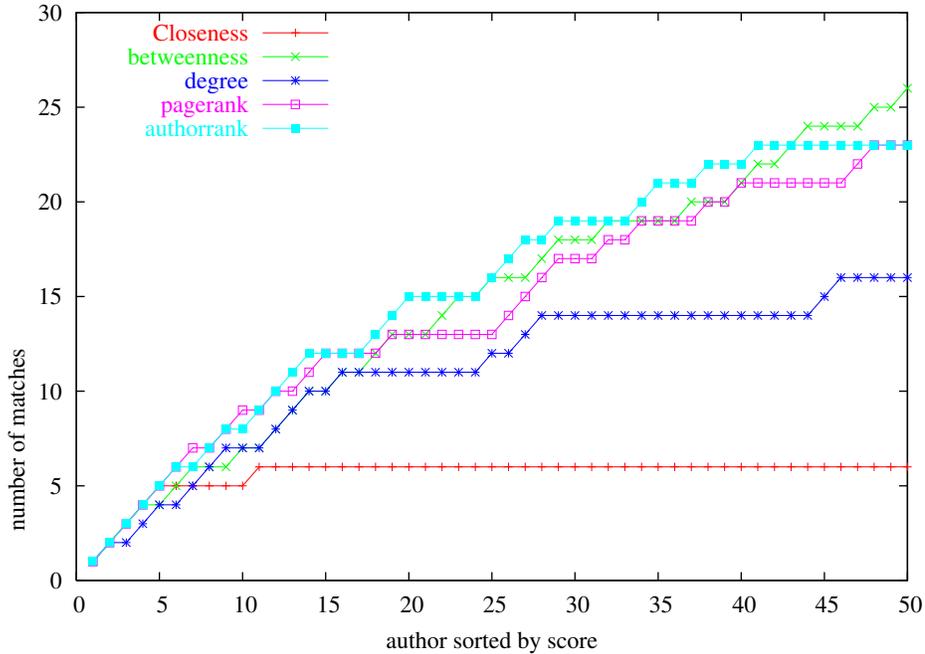}}
\caption{Ranking against JCDL program committee membership (1994-2004)}
\label{fig:committee}
\end{figure}
\end{center}
\section{Conclusions and future applications}
In this paper we investigated the co-authorship network of the DL research
community as represented in the ADL, DL and JCDL conference series.
We also presented AuthorRank, an alternative metric for ranking authors'
prestige in weighted co-authorship networks.  So what does it all
mean? What have we learned about the state of DL research 10 years after
the first DL conference?  

Our analysis paints the picture of a domain
that is in many ways still evolving the rich networks of collaboration
common in other areas of the scientific enterprise.  Our co-authorship
graphs indicate a rich tapestry of collaborations across institutional
boundaries, but demonstrate a significantly higher degree of clustering
and dispersion than one would find in other domains.  In comparison with
other co-authorship networks for related disciplines, we find the DL
research community co-authorship graph has a smaller largest component,
a larger clustering coefficient and a larger characteristic path length.
DL authors thus collaborate closely within specific clusters but restrict
their collaborations to specific groups of interest.

Do these results mean collaboration is less valued in DL research?
Of particular interest is our result demonstrating how well our
calculations of author status, i.e.~PageRank and AuthorRank, in the co-authorship
graph correspond to the JCDL program committees. Although the domain of
DLs is less well-connected than other scientific domains, the value of
collaboration still functions as an invisible hand guiding the selection
of program committees in at least one seminal DL conference.  It is thus
of vital importance that a continued emphasis be placed on collaboration
to ensure DL research will be even more of the open, diverse, but well
connected marketplace of ideas it is today.

Potentially, the presented network models have several applications.
PageRank or AuthorRank could be used as alternative metrics to evaluate research 
impacts, they can objectively guide how conference program committees are established, or to
quantitatively evaluate the prestige of conferences based on their program
committees. The weighted model has an advantage for the visualization of
a co-authorship graph, which makes it possible to emphasize important
links and truncate trivial links. Based on this idea, our colleagues
built an interactive author navigation tool \cite{996470} based on the
webdot tool of GraphViz (http://www.graphviz.org). Users can select a preferred
author (center of the graph), set a distance from the selected author,
and indicate the minimum weight necessary for links to be displayed. In
this visualization, the weight of a link plays an important role as it
allows users to identify important links.

\section{Acknowledgments}

We would like to thank Rick Luce, Linn Marks, Patrick Hochstenbach,
and Jeremy Hussell from the LANL Research Library for their support of
this work.
\bibliography{ipm-coauthor}
\end{document}